\documentclass[floatfix,aps,twocolumn,showpacs,preprintnumbers,prl]{revtex4-1}

\usepackage{times}
\usepackage{mathptmx}

\usepackage[autostyle=true]{csquotes}

\usepackage{mathtools}
\usepackage{textcomp}
\usepackage{gensymb}
\usepackage{graphicx}
\usepackage{siunitx}
\usepackage{ulem}

\usepackage[unicode=true,
 bookmarks=true,bookmarksnumbered=true,bookmarksopen=true,bookmarksopenlevel=2,
 breaklinks=false,pdfborder={0 0 1},backref=false,colorlinks=true]
 {hyperref}
\hypersetup{pdftitle={Thermally limited force microscopy on optically trapped single metallic nanoparticles},
 pdfauthor={Schnoering Gabriel},
 pdfsubject={Optical tweezer},
 pdfkeywords={Schnoering, Gabriel, ISIS, force spectroscopy, optical trapping,},
 linkcolor=black, citecolor=blue, urlcolor=blue, filecolor=blue, pdfpagelayout=OneColumn,
  pdfnewwindow=true, pdfstartview=XYZ, plainpages=false}

\usepackage[all]{hypcap}

\begin{document} 

\author{Gabriel Schnoering}
\affiliation{ISIS \& icFRC, University of Strasbourg and CNRS, 8 all\'{e}e Gaspard Monge, 67000 Strasbourg, France.}
\author{Yoseline Rosales-Cabara}
\affiliation{ISIS \& icFRC, University of Strasbourg and CNRS, 8 all\'{e}e Gaspard Monge, 67000 Strasbourg, France.}
\author{Hugo Wendehenne}
\altaffiliation[Present address: ]{Universit\"{a}t Konstanz, 78464 Konstanz, Germany}
\affiliation{ISIS \& icFRC, University of Strasbourg and CNRS, 8 all\'{e}e Gaspard Monge, 67000 Strasbourg, France.}
\author{Antoine Canaguier-Durand}
\affiliation{LKB, UPMC, ENS, and CNRS, Paris, France.}
\author{Cyriaque Genet}
\email[Email: ]{genet@unistra.fr}
\affiliation{ISIS \& icFRC, University of Strasbourg and CNRS, 8 all\'{e}e Gaspard Monge, 67000 Strasbourg, France.}

\title{Thermally limited force microscopy on optically trapped single metallic nanoparticles}

\begin{abstract}
{We propose and evaluate a new type of optical force microscope based on a standing wave optical trap. Our microscope, calibrated \textit{in-situ} and operating in a dynamic mode, is able to trap, without heating, a single metallic nanoparticle of 150 nm that acts as a highly sensitive probe for external radiation pressure. An Allan deviation-based stability analysis of the setup yields an optimal $0.1$~Hz measurement bandwidth over which the microscope is thermally limited. Over this bandwidth, and with a genuine sine-wave external drive, we demonstrate an optical force resolution down to $3~$fN in water at room temperature with a dynamical range for force detection that covers almost $2$ orders of magnitude. This resolution is reached both in the confined and freely diffusing regimes of the optical trap. In the latter, we measure $10^{-11}$ m induced displacements on the trapped nanoparticle, spatially confined within less than 25 nm along the optical axis.}
\end{abstract}

\maketitle 

\section{Introduction}

Optical traps have now become central experimental tools for measuring forces at the nanoscale with outstanding positional and force resolutions. Because of their small sizes, optically trapped objects have enabled thermally limited sensitivity, in particular in liquids \cite{MaiaNeto2015,BechingerThermoForce,ZensenAPL2016,CapassoPRL2016}. These remarkable features have been exploited in a vast variety of contexts, ranging from biology \cite{neuman2008}, non-equilibrium physics \cite{ciliberto2010}, to optomechanics \cite{marago2013}. Recent work have demonstrated how non-conservative optical force fields can be measured and spatially resolved using optically trapped dielectric particles, leading to study non-trivial Brownian type of motions \cite{wu2009,VolpeForceFields}.
In this context, using metallic nanoparticles (NPs) would offer new opportunities considering all the specific modes of actuation and control that could be implemented on metallic nano-objects through their high extinction cross sections \cite{tong2009,rings2010,ruijgrok2011,cuche2012,OrritAcousto,PhysRevLett.108.018101}. But it remains difficult to reach 3D stable optical trapping of large metallic NPs (i.e. radii larger than $50$ nm) in fluids in standard laser intensity conditions \cite{svoboda1994,hansen2005,vsiler2013}. This difficulty remains to date a strong limitation despite the potential offered by metallic NPs, in particular in the context of biophysics \cite{Moffitt2008recent}, nanosensing \cite{Halas2007NatPhot} and spectroscopy \cite{baffou2014}. 

In this paper, we propose a new setup where stable trapping conditions for single large Au NPs can be reached at low laser power. In addition, our setup offers, by construction, the possibility to inject an additional laser that can exert, in a perfectly controllable way, radiation pressure on the trapped NP, along the optical axis and independently from the restoring force at play inside the trap. Importantly, all this is achieved with negligible heating, as carefully discussed below. 

In order to qualify our setup as a force microscope, important features are implemented. The additional pushing laser is injected inside the trap in such a way as to generate an optical force field as uniform as possible. Having the strength of the external force exerted on the NP constant throughout the diffusion volume of the NP inside the trap is a central feature of our scheme that stands out from force measurements performed using total internal reflection microscopes. It indeed allows operating our setup in a dynamical mode (DM) where the pushing laser, harmonically driven at a fixed frequency, leads to a modulated force signal independent from the instantaneous position of the NP inside the trap. As well known in such conditions, the DM operation is particularly appealing from a calibration point of view since it does not request any static force calibration procedure \cite{samori2006scanning,jourdan2009,guan2015,Flyvbjerg2006,NovotnyCalib}. In our experiments, the external force is directly measured on the power spectrum density (PSD) of the motion of the trapped NP at the modulation frequency. Operating with a uniform force field, one simply demands a \textit{positional} calibration of the optical trap, obtained from the fluctuation-dissipation theorem \cite{berg2004}. This straightforward approach is clearly an advantage of our method, considering that a static force calibration is, in general, challenging to perform within the limits of stability of the setup. 

Discussing such limits is important since they play a crucial role in the determination of the force resolution level \cite{czerwinski2009,PerkinsNanoLetters2015,PhysRevLett.111.103603,schno2016}. They indeed yield the optimal (maximal) time $\tau_{\rm opt}$ over which a measurement remains thermally limited.  In our work, these limits are properly identified through a global stability analysis of the setup. The acquisition of the positional PSD of the trapped NP over $\tau_{\rm opt}$ thus directly leads to determining the smallest external force $F_{\rm opt}$ measurable on our setup. Shorter than the duration over which the NP can remained trapped, $\tau_{\rm opt}$ warns against the appealing, but misleading possibility to improve resolution by reducing the spectral noise density through an average of independent sequences of measurements extracted from acquisition time series longer that $\tau_{\rm opt}$. In fact, at the level of a single measurement performed over $\tau_{\rm opt}$, the signal associated with the external drive is superimposed to contributions coming not only from thermal fluctuations but also from other (uncharacterized) noise sources. As detailed below, the stability analysis sets for our experiments a resolution criterium at twice the thermal limit that leads us to demonstrate an external optical force resolution of $3~$fN in water at room temperature within a $\tau_{\rm opt}=10$ s measurement time.

\section{Standing wave optical trap} 

\begin{figure}[htb]
  \centering{
    \includegraphics[width=1.0\linewidth]{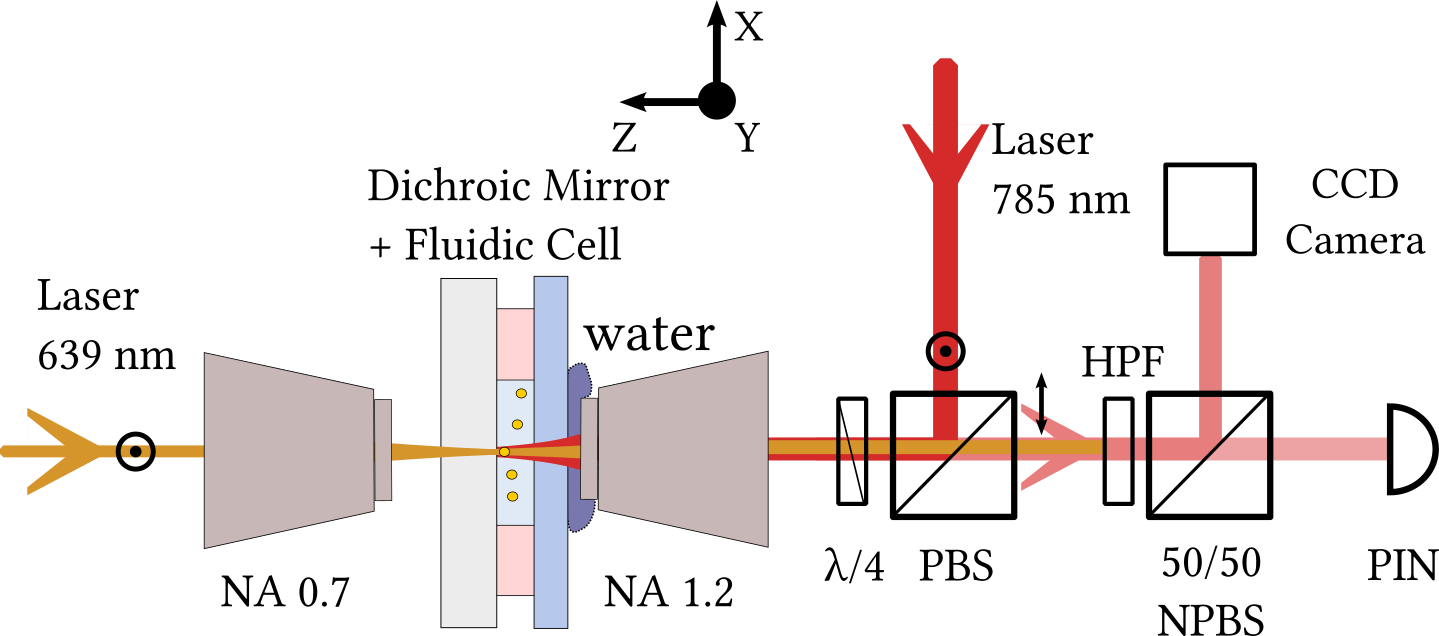}}
  \caption{Schematics of the experimental setup. Gold nanospheres (BBI) are trapped by a near IR laser (785 nm, 45 mW  Excelsior-Spectra Physics laser diode, optically isolated) with a power of 24.5 mW at the entrance of a water immersion objective (NA 1.2, $100\times$). A standing wave optical trap is formed by reflecting the laser beam on a dichroic mirror (cut-off at 700 nm). The reflected intensity $I(t)$ varies linearly with the nanosphere displacement $z(t)$ inside the trap and is collected and recorded by a PIN photodiode (Thorlabs det10A). A second beam (639 nm, 70 mW Thorlabs laser diode) of low power ($400~\si\micro$W) is injected inside the trap colinearly with the trapping beam but from behind the dichroic mirror using a dry objective (NA 0.7, 60$\times$). This second beam is not expanded and does not fill the entrance pupil of the objective. It is hence only weakly focused in front of the trapped bead and can be used to push the bead along the optical axis, minimizing any gradient force effect. To avoid any intensity signature of the modulated signal on the PIN photodiode, a high-pass filter (HPF) at 650 nm is added.}
  \label{fig1}
\end{figure}

Our optical trap configuration, schematized in Fig. \ref{fig1}, consists in focusing a trapping laser at $\lambda_T=785$ nm into a $120~\si\micro$m fluidic cell, entering with a mean intensity of 24.5 mW through a water immersion objective on top of an end-mirror. The beam {largely} overfills the objective pupil and its power is estimated to be of $\sim$10 mW at the waist. This mirror is highly reflecting at $\lambda_T$ and therefore induces a standing wave pattern inside the cell. In such a counter-propagating beam configuration, the incident and reflected scattering forces acting on the NP practically compensate each other, and can therefore be easily overcome by the gradient force induced by the focusing effect of the objective \cite{ZemanekOptLett1999}. This configuration yields a balance of forces appropriate for trapping metallic NPs of radii larger than $50$ nm, an interesting asset when aiming at measuring radiation pressure forces. Such a capacity is not found on conventional single beam traps where the scattering forces tend to push away from the waist such large metallic NPs. We have indeed checked experimentally that replacing the end-mirror by a glass slide does not allow trapping the Au NPs. This is in agreement with calculations that cannot find any stable position for such an Au NP within a propagating Gaussian beam determined from our experimental conditions of wavelength and numerical aperture \cite{brzobohaty2015}.

We monitor the instantaneous position $z(t)$ of the NP by recording the trapping laser light scattered by the NP in the forward direction and reflected back towards the PIN detector -see Fig. \ref{fig1}. In the Fourier domain, the motional dynamics of the NP is described by its power spectral density (PSD) $S_z[f]=2 |z[f]|^2$ \cite{pwelch}. In the low Reynolds number conditions of our experiments, the overdamped displacement of the NP along the optical axis of the trap obeys the spectral Langevin equation $z[f]=\chi[f]F_{\rm th}[f]$ where $\chi[f]$ is the mechanical susceptibility of the NP inside the trap and $F_{\rm th}[f]$ the Langevin force responsible for the Brownian motion of the NP. Assuming that the response of the trapped bead is harmonic, the susceptibility 
\begin{align}
\chi[f] = \frac{1}{\kappa_T-i2\pi\gamma f}
\end{align}
is Lorentzian with $\gamma = 6\pi\eta R$ the Stokes drag and $\eta\sim 10^{-3}$ Pa$\cdot~{\rm s}^{-1}$ the dynamical viscosity of water at room temperature. The stiffness of the harmonic trap $\kappa_T=2\pi \gamma f_T$ is characterized by a roll-off frequency $f_T$. 

At thermal equilibrium, the (one sided) Langevin force spectral density is given by the fluctuation-dissipation theorem (FDT) with $S_{\rm th}[f] = 4 k_{\rm B}T\gamma$. The bead dynamics is entirely driven by thermal fluctuations which have a broad Gaussian white noise spectrum. The PSD  
\begin{align}
S_z[f]=\frac{D}{\pi^2 \left(f^2+f_T^2\right)}
\label{PSDthermo}
\end{align}
thus spectrally describes the motional response of the trapped NP under the action of thermal forces with a diffusion coefficient $D=k_{\rm B}T/\gamma$. 

\begin{figure}[htb]
  \centering{
    \includegraphics[width=1.0\linewidth]{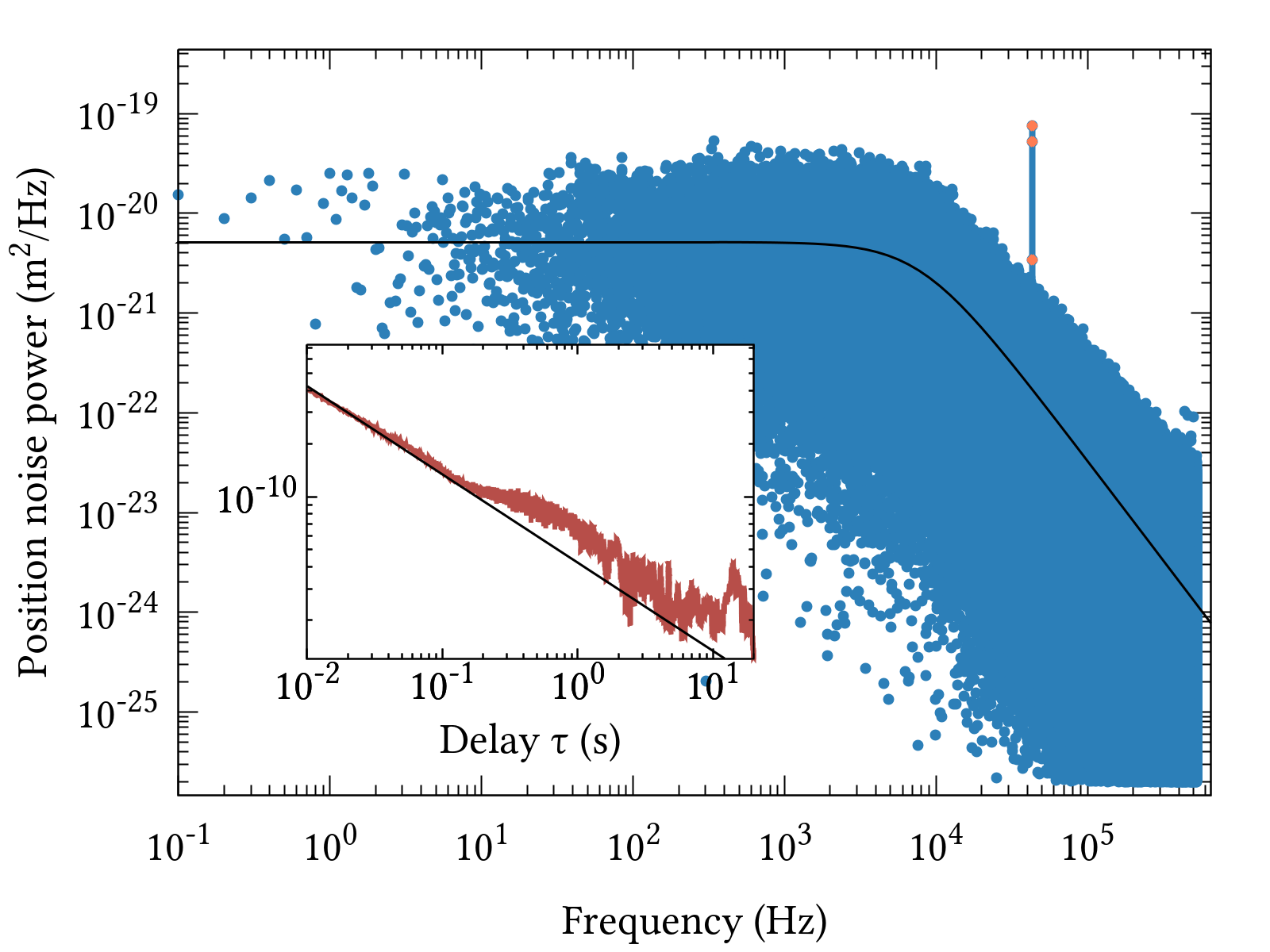}}
  \caption{One experimental (one-sided) PSD acquired over $\tau_{\rm opt}=10$ s at 1 MHz is shown in the main panel, with a roll-off frequency $f_T=8$ kHz. The corresponding Lorentzian fit is shown in black. The external drive amplitude, with a modulation ratio of $0.3$, has a strong spectral signature around $f_{0} = 43333$ Hz. The height of the peak is large and well-above the PSD variance. The spectral height of this contribution leads to a measure of the force via the trap calibration. The Allan standard deviation $\sigma_z(\tau)$ of a trapped gold nanosphere without external drive is shown in the inset. For time delays smaller than $0.1$ s, $\sigma_z(\tau)$ shows a $-1/2$ slope. This, on a PSD, is a white noise and corresponds to the plateau at low frequencies where the bead is trapped. For small frequencies (i.e. long time delays) a slight departure from the thermally driven dynamics of the bead appears. At $10~$s, it varies usually between $1.2$ and $2$ depending on the trapped Au NP and its distance to the end-mirror. }
  \label{fig2}
\end{figure}

The PSD $S_z[f]$ is determined after calibration of the intensity PSD $S_I[f]$ against the known properties of the fluid and assuming that the recorded intensity $I(t)$ is linear with the displacement $z(t)$ of the NP. This assumption holds in our experiment because the bead displacements in the trap are small and the measured PSD follows the Lorentzian fit that uses a linear restoring force.

The calibration factor is given by fitting the experimental PSD by the Lorentzian model of Eq. (\ref{PSDthermo}). The fit uses only points above 10 Hz where it is reasonable to assume that all the displacement contributions (aside from the external force peak) originate from thermal fluctuations. This provides best-fitted values for $D$ and $f_T$ and from the diffusion coefficient $D_{\rm FDT} = k_{\rm B}T/\gamma$ calculated from the FDT (assuming known temperature and viscosity), the calibration factor is simply $\beta=D_{\rm FDT}/D$ and the calibrated PSD $S_z[f] = \beta^2 S_I[f]$. For the one-sided PSD on Fig. \ref{fig2}, we extract a best-fitted roll-off frequency $f_T=8~$kHz and a calibration factor $\beta=1.2 \times 10^{-7} ~{\rm m}\cdot{\rm V^{-1}}$. 

Because the trap stiffness $\kappa_T$, directly proportional to the field intensity, depends on the viscosity $\eta(T)$, it is clear that a wrong estimation of the fluid temperature can impact the calibration of the setup and the force measurements. Optical powers at the waists of tightly focused light beams can reach significant levels (of the order of MW/cm$^2$), and metallic objects are subject to strong elevations of temperature, around 1500 K/W for Au NPs of 150 nm at $\lambda_T$ (see Supplemental Material, Sec. C). This would correspond for our experiments to an increase in temperature of ca. 40 K which gives, for water, a change in viscosity by a factor two. In order to check, and if necessary, estimate such unwanted heat contributions, we varied the trapping beam intensity for different Au NPs stably trapped, and at different mirror-waist distances. As discussed in detail in Supplemental Material, Sec. C, no deviations to linearity for trap stiffnesses as a function of the laser power were observed, suggesting a constant surrounding viscosity for all trapping laser intensities. Importantly therefore, heating of the trapped Au NP in our system, if present, has a negligible impact in calibrating the setup and measuring an external force.

\section{Test of global stability: Allan deviation analysis}

As we emphasized in the Introduction, the limited stability of the entire experimental setup in standard conditions puts an upper bound on the available measurement time. This is particularly true for our configuration where the interferometric nature of our trap makes it very sensitive to all external perturbations such as flow drift and evaporation inside the fluidic cell, vibrations, etc. Despite the fact that a single Au NP can be kept in the trap over minutes, the low frequency drift of our optofluidic system impacts the overall trapping dynamics. As a result of this uncontrolled drift, it is not possible to average a large number of measurements repeated throughout long acquisition times in order to improve the experimental sensitivity. This would misestimate the roll-off frequency.

In this context, it is important to determine the optimal data acquisition time $\tau_{\rm opt}$ beyond which the various sources of noise have drifted the entire setup out of the thermodynamic response given by Eq. (\ref{PSDthermo}) \cite{zimmermann2005,czerwinski2009}. To do so, we evaluate the Allan deviation $\sigma_z(\tau)$ for various NPs trapped for 1 minute at different mirror-waist distances. A typical $\sigma_z(\tau)$ is shown in Fig. \ref{fig2} (inset). The Allan deviation measures the standard deviation of the bead motion after it has been averaged for a time $\tau$. This analysis clearly reveals low frequency noise sources for acquisition times longer than 100 ms. Nevertheless, their contribution remains limited up to $10$ s. For all the NPs measured, differences between $\sigma_z(\tau)$ and the thermal limit at 10 s remained within 1.2 and 2 ratios. This gives an optimal time  $\tau_{\rm {opt}}\sim 10~$s where noise sources remain limited throughout all measurements. The maximal deviation from the thermal limit, a factor 2 after 10 s, also fixes a strong criterion for the minimum resolvable external force, as discussed further down. All our measurements, like the PSD displayed in Fig. \ref{fig2}, have been performed over $\tau_{\rm {opt}}\sim 10~$s, corresponding to an experimental bandwidth $\Delta f_{\rm opt}=0.1$ Hz. 

Under such conditions of optimal bandwidth, the spatial confinement offered by the optical trap at thermal equilibrium is set by the equipartition result $\delta z =\sqrt{k_{\rm B}T / \kappa_T}$. With the values extracted from the fit of the PSD in Fig. \ref{fig2}, we evaluate $\delta z = 24$ nm. Considering that the trapping position is typically located at a few (ca. $2-3$) microns from the mirror surface, such a low $\delta z$ value leads to neglect any $z$-dependent surface-induced correction to $\eta$ that can safely be taken as its bulk value of $\sim 10^{-3}$ Pa$\cdot~{\rm s}^{-1}$. Reminding that a fixed value for the viscosity is a necessary condition for the calibration procedure presented above, one understands that this $\delta z$ value is an important parameter to determine. 

\section{The force microscope}

Our force microscope, schematized on Fig. \ref{fig1}, consists in illuminating an Au NP optically trapped (in the standing wave of the trapping laser) with a second pushing laser that exerts an optical force on the NP. The wavelength $\lambda_P$ of the pushing laser is set from a Mie calculation that gives at $640$ nm a maximum in the extinction coefficient of an Au NP of radius $R=75$ nm illuminated by a plane wave. This allows for large radiation pressure effects with low laser intensities (a mean intensity of $400~\si\micro$W at the entrance of the objective).

It is absolutely crucial that the pushing laser acts on the NP independently from the restoring force at play inside the trap. To this aim, we use a dichroic end-mirror that reflects the trapping laser at $\lambda_T=785$ nm, while transmitting the pushing laser at $\lambda_P=639$ nm. The pushing laser comes from behind the mirror colinearly with respect to the optical axis of the trapping laser. It underfills a dry microscope objective (NA $0.7$, $60\times$) in such a way that, transmitted through the dichroic mirror, it is only slightly focused behind the trapped bead. Both gradient and scattering forces are at play, but we have carefully checked that the pushing laser never traps the NP and never perturbs the trap dynamics. We have confirmed numerically that at such wavelength $\lambda_P$ and using such NA, the pushing laser was not able to trap an  Au NP of radius $R=75$ nm. 

In the DM operation of the optical trap, the pushing laser power is sinusoidally modulated at a frequency $f_0$ around a mean value with $P_P=\langle P\rangle_t+P_{\rm mod}\cos (2\pi f_0 t)$. A high-pass filter above $\lambda_P$ eliminates any contamination on the scattered $\lambda_T$ signal by the modulation of the pushing laser. The overdamped dynamics of the optically trapped NP is therefore simply determined by a static (DC) force component $F_{\rm DC}$ -proportional to $\langle P\rangle_t^2$- and a modulated (AC) force component $F_{\rm AC}$ -proportional to $P_{\rm mod}^2$- both added to the thermal Langevin force, as discussed in Supplementary Material, Sec. D. This yields a spectral displacement
\begin{align}
\begin{split}
  z[f]=\chi[f] \cdot\big(F_{\rm th}[f]+F_{\rm DC}\delta[0]+\frac{F_{\rm AC}}{2}\left(\delta[f-f_0]+\delta[f+f_0]\right)\big)
\end{split}  \label{Eqz}
\end{align}
from which the contribution of the radiation pressure can be measured at the drive frequency. Experimentally, the output signal from the PIN photodiode that records $z(t)$ is sent into a low noise pre-amplifier. A high-pass filter with cut-off frequency at $0.03$ Hz removes from the signal the DC component of the force, which can then span the whole vertical resolution of the acquisition card with the bead dynamics that includes the AC force modulation. The measured signal is thereby not electronically limited. 

The pushing laser is injected in such a way that the external force field induced by the pushing laser on the NP can be considered uniform throughout the volume of the optical trap. Forces exerted on the trapped NP hence are not modulated by its Brownian diffusion. This central features gives Eq. (\ref{Eqz}) its simple structure with the external force components $(F_{\rm DC}, F_{\rm AC})$ independent from $z(t)$. Position-dependent external forces, such as found with evanescent waves involved in total internal reflection microscopy, would complicate the whole experiment. Eq. (\ref{Eqz}) also implies that the pushing laser induces no additional heating effect. We have carefully checked that this is the case, with no change in the trap stiffness observed with the external force on (see details in Supplementary Material, Sec. C).

\section{Force Measurement, Resolution, and Sensitivity }

\begin{figure}[ht]
  \centering
  \includegraphics[width=1.0\linewidth]{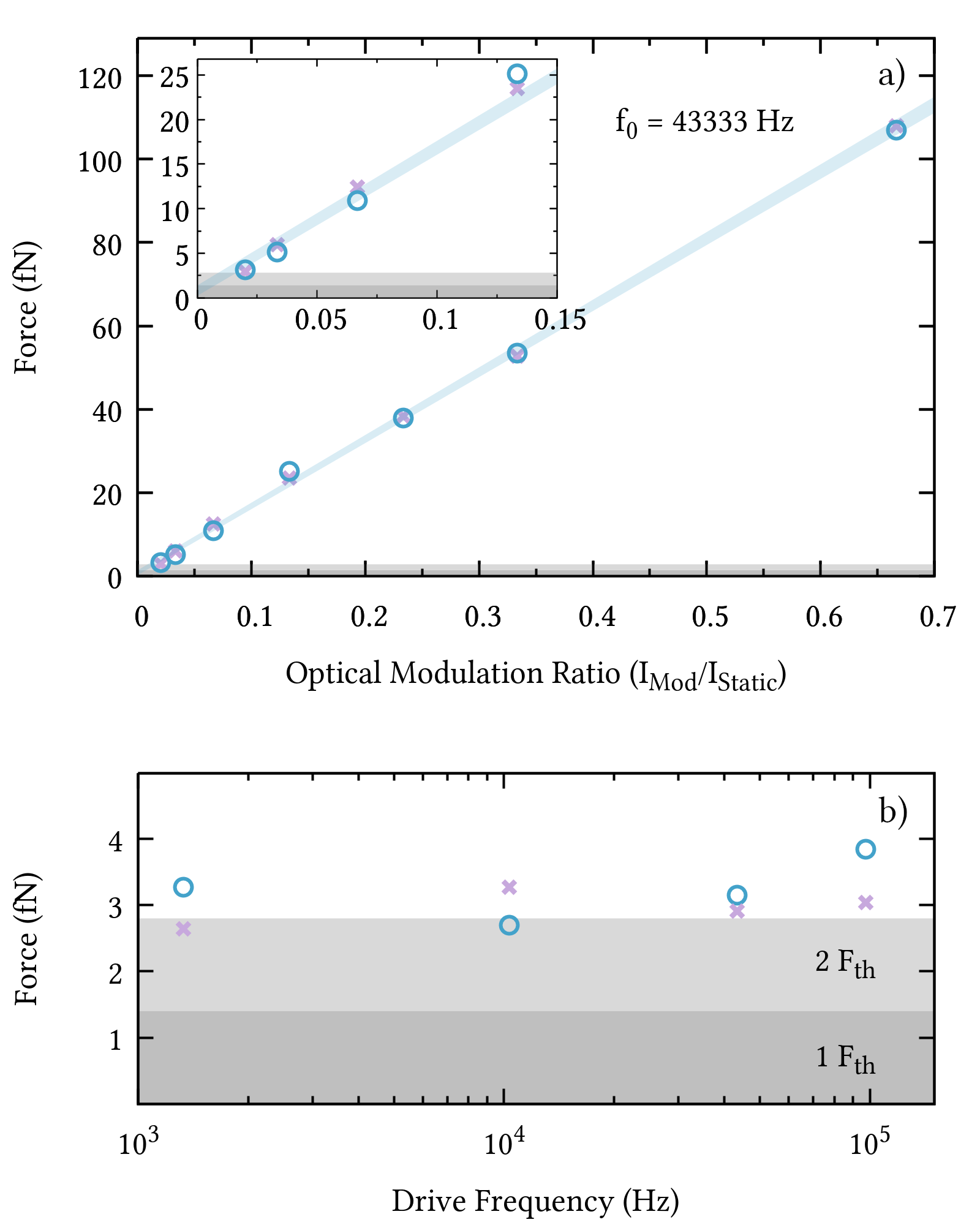}
  \caption{a) Optical forces measured at the modulation frequency $f_0 = 43333$ Hz as a function of the ratio between the modulation of the pushing beam amplitude $I_{\mathrm{Mod}}$ set by the function generator and the static pressure contribution $I_{\mathrm{Static}}$. The force is measured either by the intensity peak on the PSD (pink crosses) or with a lock-in amplifier (blue circles, see Supplementary Material, Sec. A for details). The force, as expected, varies linearly (blue shade) with the pushing laser modulation intensity. Significant external forces down to $\sim 3~$fN are measured using both the lock-in amplifier and the PSD with the relevant spectral bins populated by the external force and distributed over $\Delta f_{\rm AC}$. Note that the force resolution is independent from the static DC radiation pressure component $F_{\textrm DC}$. The smallest force measured $F_{\textrm AC}^{\rm min}$ turns out to be almost 50 times smaller than $F_{\textrm DC}$ determined to be ca. $160~$fN from the slope of the force vs. modulation ratio. 
    b) Evolution of the smallest measured force using both PSDs and lock-in measurements at different drive frequencies $f_{0}$ (going from $\sim$1 kHz to $\sim$100 kHz). The smallest measured force stays constant at ca. $3~$fN throughout the whole drive frequency range (two full orders of magnitudes) with both methods. Dark and light shaded areas represent 1 or 2 thermal forces $F_{\mathrm{th}}$ respectively at the chosen bandwidth (0.1 Hz).}
  \label{fig3}
\end{figure}

Under such sinusoidal modulation of the pushing laser, the spectral signature of the AC force component $F_{\rm AC}$ is a single resonant peak centered at the modulation frequency $f_0$ of the pushing laser. It is directly superimposed on the (one-sided) PSD $S_z^{\rm d}[f]$ of the driven trapped bead as  
\begin{align}
S_z^{\rm d}[f]=\frac{1}{\left(f^2+f_T^2\right)} \left(\frac{D}{\pi^2}+\frac{F_{\rm AC}^2}{8\pi^2\gamma^2}\delta\left(f-f_0\right)\right).
\label{PSDfull}
\end{align}

In a first series of experiments, we exert optical forces on the NP with a sufficiently strong $P_{\rm mod}$ modulation so that the peak associated with $F_{\rm AC}$ clearly emerges above the PSD noise level. Because of the electronics, the spectral density of the peak is in fact distributed over a finite frequency range $\Delta f_{\rm AC}$ which is smaller than 0.4 Hz. The peak intensity $I = \sum_i {I_i}$ is therefore determined by adding all the spectral contributions $I_i$ spread from both sides of $f_0$ over $\Delta f_{\rm AC}$ and from which the thermal contribution $S_z[f]$ (see Eq. (\ref{PSDthermo})) is subtracted. This provides an estimation of the force sensitivity (in N$/\sqrt{\rm Hz}$), given by $\sqrt{8 \pi^2 \gamma^2 I \left({f_0^2+f_T^2}\right)}$ according to Eq. (\ref{PSDfull}).
 
Taking the minimal measurable peak spectral density $I_{\rm min}$ as one standard deviation $\sigma_z[f]$ of the PSD, the sensitivity of the optical force microscope is expected to be thermally limited at $ \sqrt{8 \pi^2 \gamma^2 \sigma_z[f] \left({f_0^2+f_T^2}\right)}$ which equals $2 \sqrt{2 k_{\textrm B} T \gamma}$, using the property that for a continuous response driven by a Gaussian white noise, the standard deviation of the power spectral density equals its value ($\sigma_z[f]=S_z[f]$) \cite{berg2004}. With a Gaussian white noise, therefore, the thermal force sensitivity is only depending on the fluid properties and the radius of the NP via the Stokes drag, just like for AFM, where reducing dissipation sources is a key target in order to improve resolution levels \cite{cleland2002}. In this respect, the possibility for trapping Au NP of radius $R=75$ nm is a good compromise between the $\sqrt{R}$ dependence of the Stokes drag contribution important to reduce as much as possible in order to improve on the thermally limited force sensitivity, and the $R^3$ dependence of the absorption cross-section that determines the strength of the radiation pressure.

In practice, starting with large $F_{\rm AC}$ values, one first measures over the optimal bandwidth $\Delta{f_{\rm opt}}$ the AC force signal at $f_0$ through a high peak spectral intensity $I \gg \sigma_z[f]$. Fig. \ref{fig3} (a) gathers such force measurements obtained, with single trapped Au NP, for relatively large optical modulation ratio. We stress that all $f_0$ peak spectral intensities $S_z^{\rm d}[f]$ have been measured on a PSD (or a lock-in amplifier, as discussed in Supplementary Material, Sec. A) acquired with a bandwidth of $\Delta f_{\rm opt}$, hence at similar noise levels as the one of the PSD shown in Fig. \ref{fig2}.

Reducing the optical modulation ratio, one faces the relative increase of unavoidable noise (thermal, external, vibrations, etc) with respect to the force signal. This noise contribution is analyzed through the Allan deviation analysis. After $10$ s, this deviation (seen in the inset of Fig. \ref{fig2}) reaches, at worst, twice the thermal contribution. This sets the optimal experimental sensitivity to $2\times (2\sqrt{2 k_{\textrm B} T \gamma})=9.2 \ \textrm{fN}/\sqrt{\textrm{Hz}}$ for our experimental conditions (single Au NP, radius $75$ nm, trapped in water at room temperature) \cite{note2}. It is worth insisting that this sensitivity value is valid only for experiments shorter than $\tau_{\rm opt}$.  

Working at the optimal bandwidth $\Delta f_{\rm opt}=0.1~$Hz, we expect in these conditions a resolution of $2.9$ fN that corresponds to the minimal force value that can be measured by our microscope. All measurements of external forces below $2\times F_{\mathrm{th}}$ will be discarded, because the corresponding force signal cannot be discriminated from the noise \cite{note1}. Remarkably, as seen in the inset of Fig. \ref{fig3} (a), our system enables us to measure, directly from the $f_0$ modulation peak on the PSD, significant radiation pressure values down to 3 fN, i.e. at the level of the expected resolution. In agreement with these values, a Mie computation with field intensities estimated at the experimental limit yields a force of 4 fN exerted on the 150 nm Au sphere, a value in good agreement with our measurements.

The accuracy of these values also depends on the precise determination of the bead radius. We use manufacturer specifications and have a $8\%$ dispersion in the bead size. This variation of size induces a change in the viscosity that systematically shifts all force values, including the thermal one.  

We have also verified that the experimental resolution is independent from the modulation frequency $f_0$, as expected from a Lorentzian PSD. Fig. \ref{fig3} (b) gathers the smallest external forces measured from PSDs and with a lock-in amplifier for modulation frequencies taken below, at, and above the roll-off frequency $f_{T}$ of the optical trap. Remarkably, the values stay at around $3~$fN, i.e. ca. $2\times$ larger than the thermal limit, regardless of the modulation frequency, i.e. both in the confined $f_0<f_T$ and in the freely diffusing $f_0>f_T$ regimes of the trap.

\section{Position noise}

Measuring $S_z^{\rm d}[f]$ by selecting spectral bins on the PSD over a sufficiently narrow spectral bandwidth $\Delta f_{\rm AC}\ll f_0$, or equivalently operating through a lock-in amplifier, as described in the Supplemental Material, Sec. A, corresponds to a band-pass filter centered on the modulation frequency $f_0$. In such conditions, the position noise is given by $\delta z_{\rm min} \sim \sqrt{\Delta f\cdot S_z[f_0]}$ \cite{reynaud1990}. Contrasting with the thermal limit for force measurements, position noises therefore depend on the modulation frequency. 

\begin{figure}[htb]
  \centering
  \includegraphics[width=1.0\linewidth]{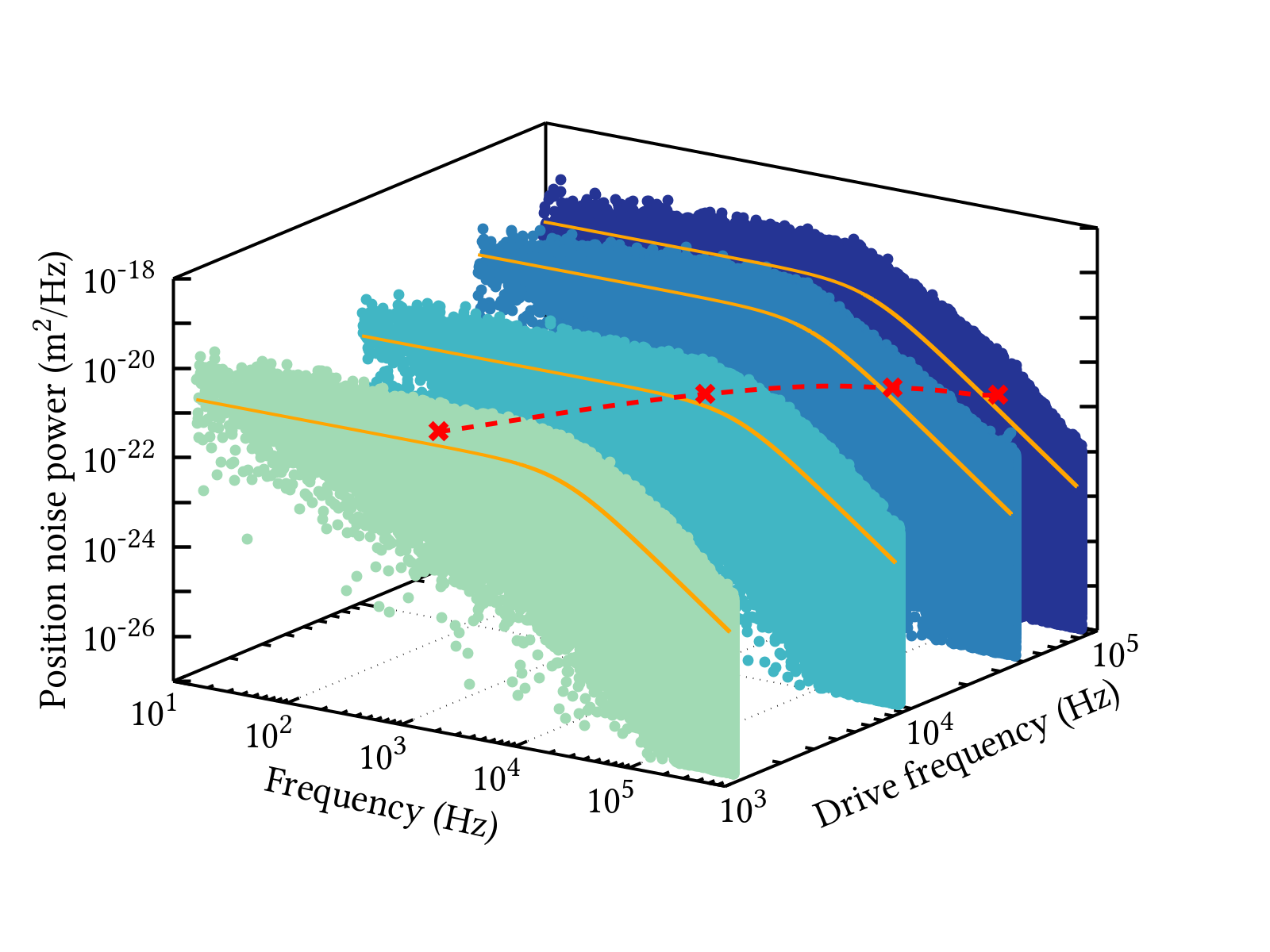}
  \caption{Isometric representation of the PSD as a function of the external drive frequency $f_{0}$ (going from $\sim$1 kHz to $\sim$100 kHz). The resolutions in position measured by the lock-in amplifier (and converted back to noise powers) are superimposed on the graph as red crosses as a function of $f_0$. The smallest measured displacement amplitude is below $1$ $\si{\angstrom}$ and diminishes even further as the drive frequency departs from the Lorentzian plateau and reaches the free brownian regime ($f^{-2}$ at high frequencies on the PSD). The fit line (in orange) on each PSD represents the thermal contribution to the displacement. All measurements have been acquired over thermally limited acquisition time $\tau_{\rm opt}=10$ s. Experiments were performed with different nano-spheres and at different distances from the mirror. This results in different trap stiffnesses for the different acquired series. The roll-off frequencies are of 17 kHz for $f_0 = 1331$ Hz, 18.5 kHz for $f_0 = 10331$ Hz, 8 kHz for $f_0 = 43333$ Hz and 7.2 kHz for $f_0 = 97579$ Hz.}
  \label{fig4}
\end{figure}

Displacements associated with the smallest measured external forces are displayed on Fig. \ref{fig4}. As clearly seen, they lie within the PSD noise levels, separated only by a factor ca. $2$ from the Lorentzian fits. The $f_0$ dependence yields $\delta z_{\rm min}$ that rapidly decrease with $f^{2}$ as soon as the free Brownian regime is dominant for $f_0>f_{T}$. Furthermore, the large $f_{T}$ values provide sub-$\si{\angstrom}$ levels of resolution in position for all drive frequencies $f_0$ and displacements of $10^{-11}$ m reached at $f_0 \sim 100~$kHz.

\section{Conclusion}

The careful assessment of the conditions of stability of our experiment through the Allan deviation analysis validates our setup as a high resolution optical force microscope.  Over thermally limited $0.1~$Hz bandwidths, we have been able to consistently measure radiation pressure forces down to $\sim 3$ fN. This result should also be appreciated in relation with a dynamical range $F_{\rm AC}/F_{\rm DC}$ of ca. $2$ orders of magnitudes. This range, together with the capacity to reach fN force resolution levels in water, at room temperature on relatively short acquisition times, in the absence of any induced heating, is particularly important when aiming at studying new types of optical force fields \cite{ruffner2012}, in particular in the context of surface plasmon optics \cite{rodriguez2015}, optical spin-orbit interactions \cite{sukhov2015} and chiral optical forces \cite{canaguier2013,cameron2014,ding2014}. 

The concomitant sub-$\si{\angstrom}$ resolution in displacements offered by our setup also opens new possibilities in the context of short-distance forces, such as Casimir-like interactions \cite{MaiaNeto2015} or optical binding effects \cite{dholakia2010}, where adjustable roll-off frequencies allow tuning the diffusion volume of the trapped NP, and thereby giving a capacity of localization on nm scales. This capacity could be important for resolving non-linear force signals, such as found at the level of self-organized supramolecular assemblies in mechanochemistry \cite{mechanochem2016}. In this context, the reliability and resolution provided by our force microscope could help in exploring connections between optical force and chemical signals.

\section{Acknowledgments}

 This work was supported in part by the ANR Equipex ``Union'' (ANR-10-EQPX-52-01), the Labex NIE projects
 (ANR-11-LABX-0058-NIE) and USIAS within the Investissement dÕAvenir program ANR-10-IDEX- 0002-02
 program ANR-10-IDEX-0002.

\section{Appendix A: Lock-In Detection Method}

\begin{figure}[htb]
  \centering
    \includegraphics[width=0.9\linewidth]{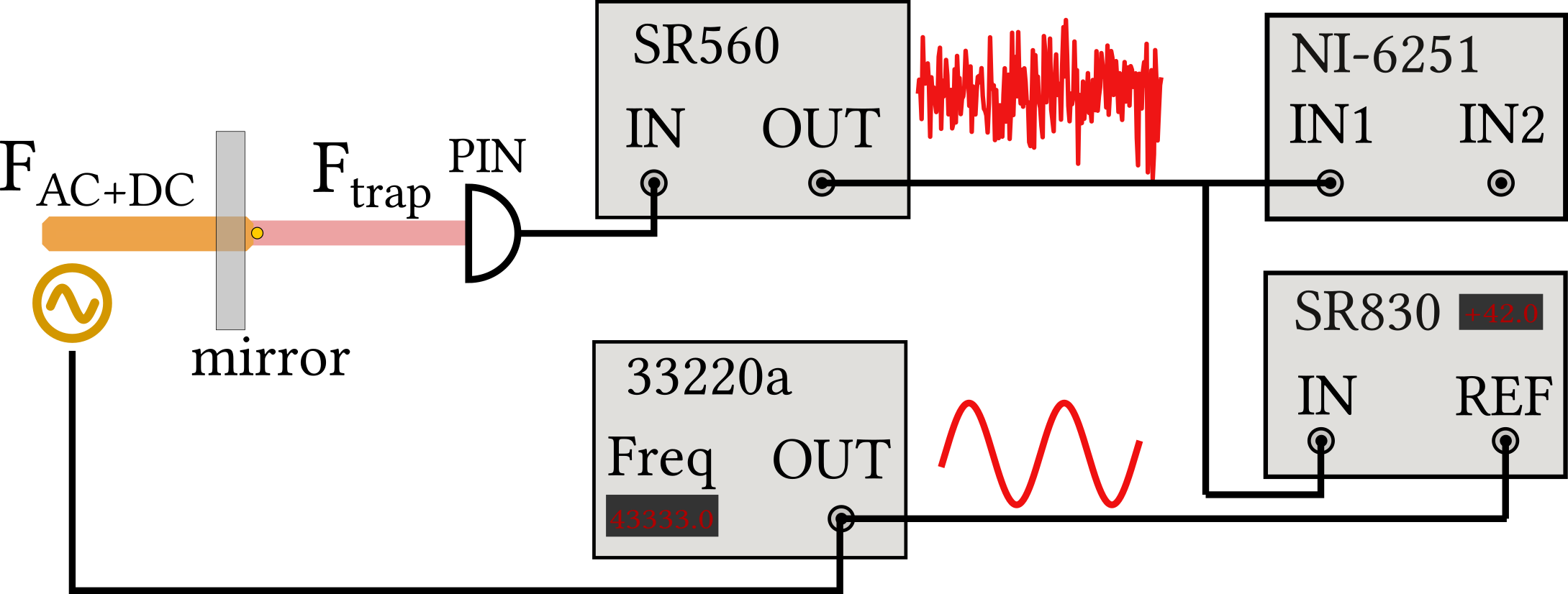}
  \caption{Schematized structure of the experimental setup used to measure $S_z^{\rm d}[f_0]$ -Eq. (\ref{PSDfull}) above. The voltage recorded by the photodiode (PIN) is sent to a low-noise pre-amplifier (Stanford Research, SR560). The signal is filtered to remove its DC component with a high-pass filter at 0.03 Hz. It is then sent to both a 16-bit acquisition card (National Instruments, NI-6251) and a lock-in amplifier (SR830). A function generator (Agilent 33220a) produces a sinusoidal output of amplitude and frequency controlled by a computer. The sine function is sent simultaneously to the pushing laser controller and the lock-in amplifier as the reference.}
  \label{fig:electronics}
\end{figure}

When measuring forces, in particular via the determination of the position noise $S_z^{\rm d}[f_0]$ (see Eq. (\ref{PSDfull}) above), it becomes convenient to resort to a time-domain measurement such as a lock-in detection method \cite{GaussianKouh}. Experimentally, $S_z^{\rm d}[f_0]$ is measured from the intensity signal passing through a low-noise pre-amplifier which signal is then sent to both an acquisition card and a lock-in amplifier, as described in Fig. \ref{fig:electronics}.

The lock-in amplifier immediately yields the spectral power signal $S_z^{\rm d}[f_0]$. However, its estimation from the complete bead displacement PSD can be computed off-line directly from the recorded intensity time-trace saved on a computer. The lock-in method becomes a viable alternative only when its output signal is calibrated.

This calibration is not straightforward since the lock-in amplifier mixes, in the time domain, both the contributions of the thermal fluctuations and of the external AC drive that cannot therefore be isolated from one another. But the lock-in calibration can be performed through the results obtained with the \textit{self-calibrated} PSD approach at high modulation amplitudes -modulation ratios larger than 0.1 in Fig. \ref{fig3} (a) in the main text -where all noise contributions to the signal can be neglected. Both methods (PSD and lock-in amplitude) provide values proportional to $S_z^{\rm d}[f_0]$. In this regime of strong drive, the linearity of the lock-in output signal is calibrated to the linearity of the power spectral intensity peak at $f_0$ measured on the PSD and converted, as $F_{\rm AC}$, in Newton, as discussed above. The experimental output signals are displayed in Fig. \ref{fig3} in the main text and superimposed to those obtained by the PSD method. The calibrated lock-in may relieves the constrain of identifying spectral bins populated by the external excitation, but in our case, both approaches are strictly equivalent. Spectral bins are well identified, even for low external drives. This equivalence clearly appears on Fig. \ref{fig3} in the main text where PSD values match well with lock-in values.

When the modulation amplitude of the pushing laser is reduced, we still measure a proportional force amplitude up to a few fN, as observed in the inset of Fig. \ref{fig3} (a) in the main text. In these conditions, it becomes possible to perform measurements down to the $2\times F_{\rm th}$ limit, below which noise sources dominate over the external force. Fig. 3 (a) in the main text shows that the smallest significant external force we measure using the lock-in amplifier is at the $3$ fN level, i.e. similar than the level reached directly on the PSD. 

\section{Appendix B: Force measurements at different modulation frequencies}

\begin{figure*}[htb]
  \centering
  \includegraphics[width=0.9\linewidth]{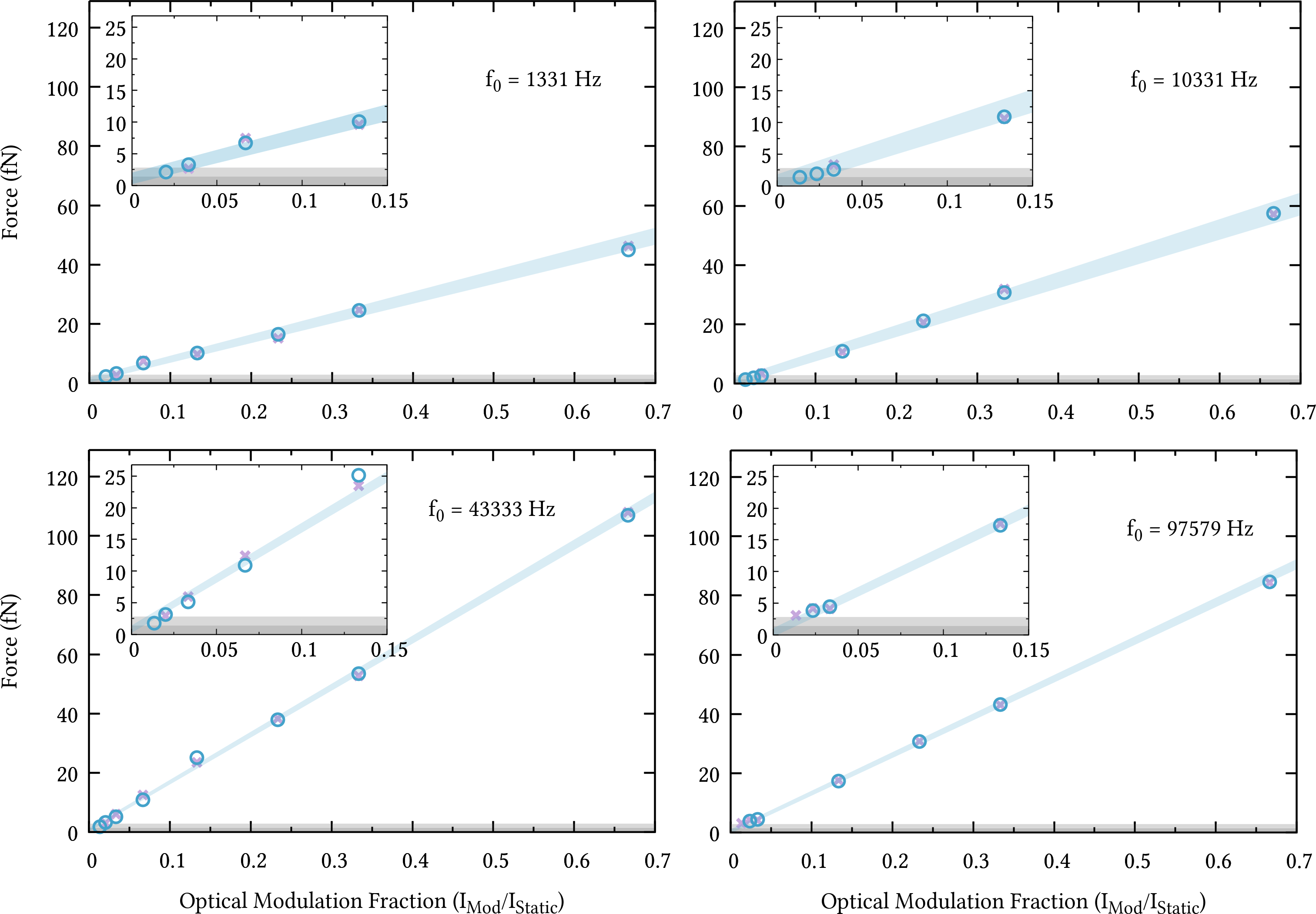}
  \caption{Measured external forces driven at $f_0$ exerted on optically trapped single Au NPs of 150 nm at frequencies: 1331, 10331, 43333 and 97579 Hz. The amplitude of the drive is modulated with respect to the mean intensity of the optical beam ($\lambda = 639$ nm). Pink crosses are measured external forces from the PSDs and blue circles represent forces measured from the lock-in approach after its response is linearly calibrated from strong external drives. Each value has a bandwidth $\Delta{f} = 0.1\ \mathrm{Hz}$. The linearity of measured forces with modulation amplitudes is quantified throught an uncertainty interval (a deviation to linearity to $\pm 1\sigma$) on the whole measurement series. Dark and light shaded areas represent 1 or 2 thermal forces $F_{\mathrm{th}}$ respectively at the chosen bandwidth (0.1 Hz).}
  \label{fig:drivesummary}
\end{figure*}

The external force is measured at different modulation (driving) frequencies $f_0$ corresponding to confined ($f_0<f_T$) or freely ($f_0>f_T$) diffusing Brownian motion within the trap. For an overdamped Brownian particle solely driven by thermal fluctuations, the measured external force, and in our case, the minimal measurable external force, is expected to remain constant regardless of the modulation frequency. This is verified experimentally with modulation frequencies $f_0$ equal to 1331, 10331, 43333 and 97579 Hz, spanning two orders of magnitude and crossing through the trap roll-off frequency. The results are shown in Fig. \ref{fig:drivesummary}.

For these experiments, an external driving force is applied on single 150 nm Au NPs trapped a few $\si\micro$m away from the mirror. The amplitude of the drive is modulated with respect to the mean intensity of the optical pushing beam ($\lambda = 639$ nm). Pink crosses indicate external force values extracted from the PSDs and blue circles represent forces measured by the lock-in amplifier after its response is linearly calibrated for strong external drives (see Sec. A above). We stress that each value has been recorded over the same bandwidth $\Delta{f} = 0.1\ \mathrm{Hz}$. The linearity of measured forces with modulation amplitudes is apparent and quantified by an uncertainty interval (a deviation to linearity to $\pm 1\sigma$) on the whole measurement series in blue shades. Grey areas correspond to thermal force noise floors at 1 and 2 standard deviations for dark and light regions respectively.

All recorded series display significant external forces that are measured above the stringent resolution criterion of $2 \times F_{\rm th}$ that we selected, i.e. $2.9$ fN. Remarkably, while the bandwidth was chosen to account for a worst-case stability scenario for a single NP, the good linearity of the overall series (taking up to a few minutes) suggests that longer acquisition times do not deviate much, for good series, from the Allan variance at 10 s. This could allow for even shorter bandwidth when considering single measurements, although such a possibility must be asserted through a long-time Allan stability analysis of the system. This however turns difficult to implement with our colloidal suspensions and our acquisition card.

\section{Appendix C: Heating effects}

Heating effects due to temperature elevation at the surface of the metallic sphere under laser irradiation. The rise in temperature can be estimated considering the radius $R$ and the absorption cross-section $\sigma_{abs}$ of the sphere, as well as the water thermal conductivity $k_s$ and the irradiance $I$: \cite{BaffouHeat}
\begin{align}
\Delta{T}=\frac{\sigma_{abs} I}{4 \pi k_s R}. \label{baffou}
\end{align}

Such effects have been measured experimentally, by looking, for instance, at trap stiffness variations \cite{Seol:06}, shifts of the localized plasmon resonances \cite{Andres-Arroyo:15} or thermal damaging of supporting membranes \cite{OddershedeNano}. These experiments provide a value of thermal elevation of ca. 500 K/W for Au spheres of 100 nm under a Gaussian illumination at a 1064 nm wavelength. These values are in relatively good agreement with Eq. (\ref{baffou}).

Adjusting this value for our 150 nm spheres illuminated at 785 nm provides temperature elevations over 1500 K/W, which corresponds to an increase of 40 K with our $\sim$25 mW laser. The viscosity of water $\eta = \eta(T)$ being strongly dependent on temperature variations, this increase is expected to lead to a factor 2 change in the viscosity -$\eta$(T = 300 K) = 0.85 and $\eta$(T = 340 K) = 0.42- that necessarily would alter our external force estimation by the same factor.  

\begin{figure}[htb]
  \centering
  \includegraphics[width=1.0\linewidth]{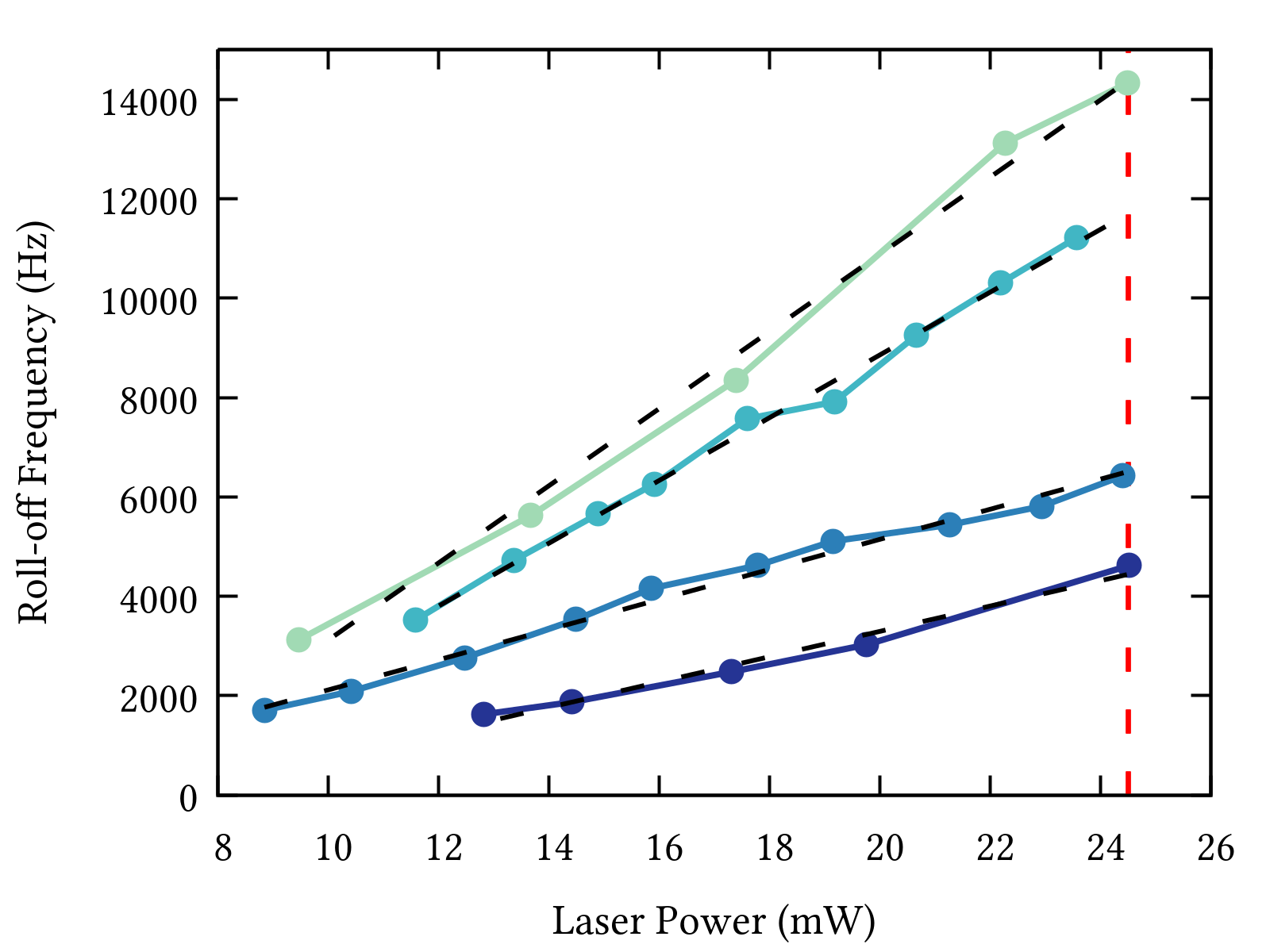}
  \caption{Evolution of the trap stiffnesses with incident trapping laser powers for 4 different Au spheres of 150 nm and at different stable trapping positions. Regardless of the mirror-waist distances, the trap stiffnesses, while different, increase linearly with increasing laser power. The behavior is expected when the drag term $\gamma$ remains constant. This implies that the viscosity of the fluid, and therefore its temperature, do not evolve sufficiently to be detected within the trapping laser power range explored here. This verifies that the possible heating of the trapped NPs remains small enough as to not induce any significant variations of the surrounding viscosity. The second top line is offset by -1000 Hz to ease readability. The red dashed line (24.5 mW) corresponds to the trapping laser intensity which is used in our experiments.}
  \label{fig:stiffpower}
\end{figure}

But such a change is not observed in our experiments. We carefully checked this by varying, through a rotating optical density, the trapping laser power for single Au NPs (150 nm) trapped at different distances from the mirror. As clearly seen in Fig. \ref{fig:stiffpower}, the roll-off frequencies for all trapping conditions follow a linear behavior. This behavior is expected in the absence of heating effects, considering that the trapping roll-off frequency $f_T$ is directly proportional to the trapping laser intensity. Looking at the trap stiffness $\kappa_T = 12\pi^2 \eta(T) R f_T$, this linear dependence shows that the viscosity $\eta(T)$ of water inside the trap must remain constant throughout the variations of intensity. We can therefore conclude that no significant heating effect is at play on the dynamics of the trapped object in our experiments.

Trapping of spherical Au NP in a standing wave optical trap is characterized by complex patterns of stability regions \cite{SILER2013}. In our experiments, the absence of heating leads to infer that the NP is trapped outside field intensity maxima of the standing wave. Our system thus appears analogous to a cage with optical walls along the optical axis, preventing the NP to cross anti-nodes of the interference pattern, while being confined transversally by the residual gradient contributions of the trapping beam.

Finally, no changes were observed in the trap stiffness with the external DC force present. The external pushing field is a few orders of magnitude weaker than the trapping beam and does not induce changes in the fluid properties despite its higher absorption cross-section at the pushing laser wavelength.

\section{Appendix D: Sinusoidal forcing}

\begin{figure}[htb]
  \centering
  \includegraphics[width=0.8\linewidth]{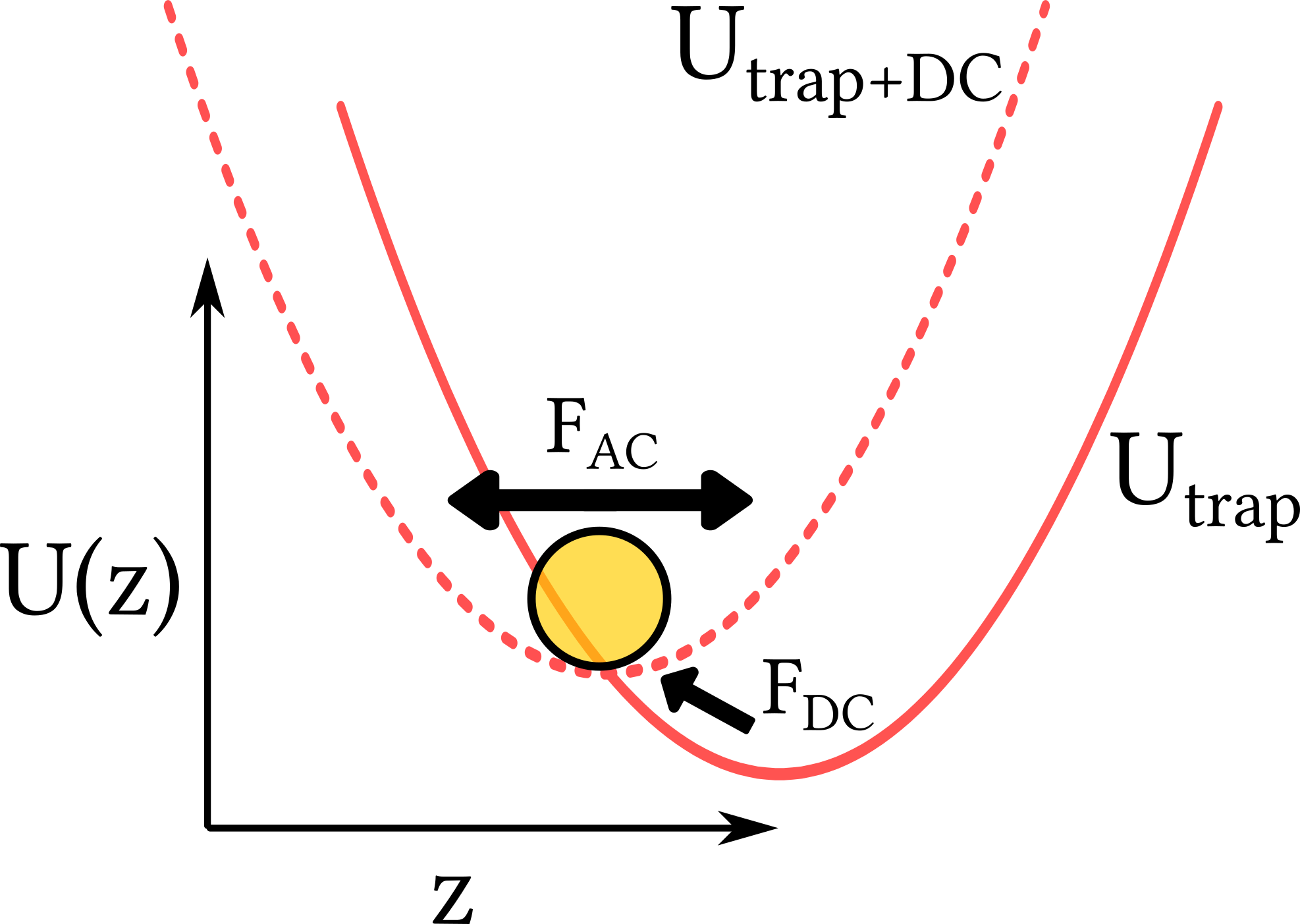}
  \caption{Simplified energy landscape of the trapped Au nano-sphere under an oscillating external forcing. This contribution consists of a sinusoidal modulation of the pushing laser intensity around a constant value. As a consequence, the resulting force applied on the sphere consists in two terms: $F_{\rm DC}$ and $F_{\rm AC}$. The constant contribution ($F_{\rm DC}$) shifts the equilibrium position of the trapped object (solid line) into a new effective harmonic potential (dashed line) and equilibrium position. The oscillating term ($F_{\rm AC}$) at $f_0$ will sinusoidally push and pull the confined metallic sphere with respect to this new equilibrium position.}
  \label{fig:harmdrive}
\end{figure}

In our experiments, external forces applied on the optically trapped metallic nanospheres are generated by sinusoidal forcing. This means that, relatively to a mean position set by the static (DC) contribution of the forcing, the sinusoidal modulation effectively pushes and pulls the particle every half period. This is achieved experimentally by modulating the intensity of the pushing laser sinusoidally around a mean value. The resulting beam intensity thus includes a static (DC) and a dynamic (AC) components. The static contribution $F_{\rm DC}$ will push the sphere and displace it with respect to the initial trapping potential. The Gaussian dynamics of the sphere displacements is preserved but the equilibrium position is shifted along the optical axis. This amounts to defining a new (shifted) effective attractive potential for the bead motion, drawn as a dashed line in. Fig. \ref{fig:harmdrive}. The dynamic contribution $F_{\rm AC}$ to the force moves the particle back and forth sinusoidally in this new effective potential. 

It is worth insisting that this description matches perfectly the dynamics observed experimentally. It is clear on Fig. \ref{fig2} in the main text that even for high external modulation strengths, no harmonics to the drive frequency $f_0$ are observed in the spectral signatures of the PSD. This contrasts with previous work where the periodic excitation is created by a chopper in order to modulate the signal on/off \cite{ZensenAPL2016,CapassoPRL2016}. In this case, odd harmonics to $f_0$ are necessarily present in the PSD and the measured $S_z^{\rm d}[f]$ amplitude has to be corrected from those, in order to account for the square nature of the driving signal.

\bibliography{biblio-forces}

\begin{thebibliography}{53}
\bibitem [{not({\natexlab{b}})}]{note1}%
  \BibitemOpen
  \href@noop {} {\bibfield  {journal} {\bibinfo  {journal} {Note that averaging
  the PSD necessarily reduces the frequency bandwidth because each signal
  series involved in the averaging is measured on a shorter time. As a
  consequence of this reduction in bandwidth, the width of the peak associated
  with the external modulated force is increased and its height reduced. This
  implies that the improvement on the PSD variance is lost when measuring the
  peak at a smaller bandwidth. In fact, for a system without additional
  external noise sources, both approaches (full signal vs. extracted series)
  should give the same measurement on the external displacement. However in our
  configuration, averaging the spectral density can lead to increasing errors
  in the measured force. As it is seen on the Allan deviation, one position
  measurement performed over $1~$s is already not solely thermally driven.
  Averaging our signal acquired over $10~$s as $10\times 1~$s will necessarily
  add the non-thermal noise observed at $1~$s. The ratio between the signal
  measured and the thermal floor due to other noise contributions being
  comparable between $1~$ and $10~$s, errors would simply be $10\times$ larger
  on the averaged position dynamics than for a single measurement of $10~$s. In
  our case therefore, averaging the PSD cannot provide any improvement in the
  force measurement.}\ } ({\natexlab{b}})}\BibitemShut {NoStop}%
\end{thebibliography}

\end{document}